\documentclass[conference]{IEEEtran}
\ifCLASSINFOpdf
\else
\fi

\usepackage{array}
\newcolumntype{P}[1]{>{\centering\arraybackslash}p{#1}}
\newcolumntype{M}[1]{>{\centering\arraybackslash}m{#1}}

\usepackage{amsmath,amsthm,amssymb,graphicx,tikz, bbold}
\usepackage[bookmarks=false]{hyperref}

\usepackage{amsfonts}
\renewcommand{\vec}{\mathbf}

\begin{document}
%
\title{A Noncoherent Space-Time Code from\\ Quantum Error Correction}




%
\author{\IEEEauthorblockN{S. Andrew Lanham\IEEEauthorrefmark{1}\IEEEauthorrefmark{2},
Travis C. Cuvelier\IEEEauthorrefmark{1}\IEEEauthorrefmark{2},
Corey Ostrove\IEEEauthorrefmark{3}, 
Brian La Cour\IEEEauthorrefmark{2},
Granville Ott\IEEEauthorrefmark{2}, and 
Robert Heath Jr. \IEEEauthorrefmark{1}}
\IEEEauthorblockA{\IEEEauthorrefmark{1}Department of Electrical and Computer Engineering} \IEEEauthorblockA{\IEEEauthorrefmark{2}Applied Research Laboratories} \IEEEauthorblockA{\IEEEauthorrefmark{3}Department of Physics\\ The University of Texas at Austin, Austin, TX 78712 USA\\
Emails: sa\_lanham@utexas.edu, tcuvelier@utexas.edu, costrove@utexas.edu, blacour@arlut.utexas.edu\\ ott@arlut.utexas.edu, rheath@utexas.edu}}


\maketitle

\begin{abstract}
In this work, we develop a space-time block code for noncoherent communication using techniques from the field of quantum error correction. We decompose the multiple-input multiple-output (MIMO) channel into operators from quantum mechanics, and design a non-coherent space time code using the quantum stabilizer formalism. We derive an optimal decoder, and analyze the former through a quantum mechanical lens. We compare our approach to a comparable coherent approach and a noncoherent differential approach, achieving comparable or better performance. 
\end{abstract}


%
\IEEEpeerreviewmaketitle

\section{Introduction}

Noncoherent communication is characterized by a lack of channel knowledge at both transmitter and receiver. This scenario is assumed when the propagation channel changes rapidly, or with frequency hopping waveforms, where the training required for coherent communication takes too much overhead. In this paper, we develop a code for noncoherent communication based on quantum stabilizer codes \cite{gottesman1997stabilizer}. We begin with a decomposition of the channel into elements of the Pauli group, a well-known matrix basis from quantum mechanics. We view our transmitted information as a quantum state and construct a code designed to mitigate the effects of the channel at infinite SNR. We then derive an optimal decoding rule for the noisy case and conclude with numerical simulations that benchmark our code against  coherent and differential schemes of comparable rates. 

The theory of quantum error correction (QEC) defines conditions allowing for the correction of a broad class of channel errors \cite{calderbank1996good} \cite{knill2001scheme}. Many families of QEC codes were developed by directly extending classical error correcting codes. For example, Calderbank-Shor-Steane (CSS) codes generalize self-dual classical codes to the quantum setting \cite{calderbank1998quantum}. A CSS construction based on LDPC codes was developed for protecting large blocks of qubits against noise \cite{mackay2004sparse}. Our work looks in the other direction, in line with \cite{eldar2002quantum}, using quantum mechanical concepts to inspire algorithms for classical communication in the noncoherent setting.

Estimation of the capacity of the noncoherent channel was derived for particular cases in \cite{marzetta_capacity_1999} and extended in \cite{zheng2002communication}. The capacity achieving code construction performs sphere packing on the Grassmann manifold, leading to codes using  Grassmannian packings \cite{gohary2009noncoherent} \cite{kammoun2003new}. Like these codes, other space-time codes based on frame theory have been considered \cite{heath2002linear}. A differential encoding based on matrix groups was proposed in \cite{differential}. The approach implicitly performs channel estimation, although the estimate is updated using only data from two coherence intervals \cite{differential}. Differential coding was extended to matrix families that are not groups in \cite{shokrollahi2001representation} and \cite{hassibi2002cayley}. 

The approaches in \cite{ashikhmin2010grassmannian} appear to be the most relevant to this work. A subspace code, namely a Grassmannian packing for noncoherent communication, was derived using formalism from the Pauli group and quantum stabilizer codes. While we use both techniques related to quantum stabilizer coding and Grassmannian packings, our approach to the coding problem is novel. We view the noncoherent channel for a specific MIMO architecture through the lens of quantum errors.  We consider the transmitted symbol as a quantum state and design a code to reconstruct that state at the receiver. Our use of Grassmannian frames occurs in a completely different dimension than in \cite{ashikhmin2010grassmannian} and \cite{zheng2002communication} and is derived from the problem of finding quantum states that are maximally separated with respect the fidelity metric.

The organization of the paper is as follows. In Section II we introduce the necessary background in quantum mechanics and review stabilizer codes. In Section III we outline the system model. In Section IV we decompose the wireless channel into quantum operators, and continue by introducing a noncoherent space-time code based on the stabilizer family of codes. We derive the optimal decoder, and analyze the former in both the classical and quantum mechanical lights. Section V characterizes the performance of the code in a Rayleigh-fading environment and demonstrates competitive performance against other approaches. 

\textbf{Notation:}
We use bold lower case letters $\mathbf{a}$ to denote column vectors, and bold upper case letters $\mathbf{A}$ to denote matrices. We use non-bold letters to denote scalars. We denote the element in the $i^{\rm th}$ row and $k^{\rm th}$ column of a matrix $\mathbf{A}$ by $[\mathbf{A}]_{i,k}$. In general we denote the $k\times k$ identity matrix by $\mathbf{I}_k$. The $2\times 2$ identity matrix is used so often that we drop the subscript, i.e. $\mathbf{I}_{2} = \mathbf{I}$. We use $\text{tr}(\mathbf{A})$ to denote the trace, $\text{det}(\mathbf{A})$ the determinant, $\mathbf{A}^T$ the transpose, and $\mathbf{A}^*$ the conjugate transpose. For positive semi-definite matrices, $\mathbf{A}^{\frac{1}{2}}$ denotes the matrix square root. We use $|a|$ to denote the absolute value of a scalar. We use $\mathbb{E}(\cdot)$ to denote expectation. We use
$\otimes$ to denote the tensor product when acting on vector spaces (i.e. $\mathbb{C}^2\otimes\mathbb{C}^2$) and to denote the Kronecker product when acting on vectors or matrices. We use $\mathcal{N}_C(\boldsymbol{\mu},\boldsymbol{\Sigma})$ to denote a complex circularly symmetric normal distribution with mean $\boldsymbol{\mu}$ and covariance $\boldsymbol{\Sigma}$. If $A = c B$ where $c >0$, we write $A\propto B$.
\section{Just Enough Quantum Mechanics}
In this section we review some material from quantum information processing. We begin by introducing general systems of \textit{qubits}, or quantum bits, that are the natural generalization of a bit in quantum computing. We continue with a definition of an important measure of distance between quantum states. Finally, we describe stabilizer codes, a powerful class of quantum error correcting codes. 
\subsection{Quantum states, measurements, and fidelity}\label{ss:intro2qm}
A qubit represents the state of a two-level system, such as the polarization of a photon, and is the most elementary example of quantum state. A qubit is represented as a state vector $\mathbf{q} = [\alpha,\beta]^T \in \mathbb{C}^2$ with $\mathbf{q}^* \mathbf{q} = |\alpha|^2+|\beta|^2 = 1$ by convention. Equipping $\mathbb{C}^2$ with the standard inner product  $\langle \mathbf{q}, \mathbf{p}\rangle  = \mathbf{q}^* \mathbf{p}$ leads us to define a qubit state as an element of a two-dimensional complex Hilbert space. In discrete time, the evolution of a closed quantum system is unitary. That is to say, $\mathbf{q}_{n+1} = \mathbf{U} \mathbf{q}_{n}$, where $\mathbf{U}\in \mathbb{C}^{2\times2}$ with $\mathbf{U}^* \mathbf{U} = \mathbf{I}$. The concept of \textit{applying} unitary operators (which is possible to good approximation) comes up often in quantum computing and will be used in this paper \cite{mikeike}. 

In general, the act of measuring, or observing, a quantum state causes the system to change. An important class of quantum measurements are \textit{projective measurements}. Projective measurements are defined in terms of an \textit{observable}, a Hermitian operator $\mathbf{M}$ on the state space of the system of interest \cite{mikeike}. Let $m$ denote an eigenvalue of $\mathbf{M}$ and let $\mathbf{P}_m$ be the projector onto the $m$ eigenspace. The observable $\mathbf{M}$ can thus be orthogonally diagonalized as
\begin{align}\label{eq:obs}
\mathbf{M} = \sum\limits_{m} m \mathbf{P}_m \; ,
\end{align}
where $\mathbf{P}_n \mathbf{P}_m = \mathbf{0}$ when $n\neq m$. The outcome of ``measuring the observable $\mathbf{M}$" is an eigenvalue $m$. If $\boldsymbol{\psi}$ is a quantum state, then the probability of measuring $m$ is given by $p(m) = \boldsymbol{\psi}^* \mathbf{P}_m \boldsymbol{\psi}$ \cite{mikeike}. Given that the outcome $m$ occurs, the system after measurement collapses to the state $\mathbf{P}_m \boldsymbol{\psi}/\sqrt{p(m)}$ \cite{mikeike}. It turns out that projective measurements, coupled with unitary evolution, fully describe general quantum measurements \cite{mikeike}. 

A notable feature of quantum measurement is that the \textit{global phase} of a state is not observable. If $\mathbf{x} = [\alpha,\beta]^T$ and $\mathbf{y} = e^{j\theta} [\alpha,\beta]^T$, then, for a  measurement in all possible bases, the distributions of outcomes for $\mathbf{x}$ and $\mathbf{y}$ are the same. For this reason, one often works with \textit{density matrices}. A state $\mathbf{q}$ can be represented by its density matrix $\mathbf{Q} = \mathbf{q} \mathbf{q}^*$. All of our previous formalism can be represented analogously. A state $\mathbf{Q}$ that evolves by the unitary $\mathbf{U}$ becomes the state $\mathbf{UQU}^*$. The Born rule for projective measurements says that a state $\mathbf{Q}$ evolves to $\mathbf{P}_m \mathbf{Q} \mathbf{P}_m/p(m)$ with probability $p(m) = \mathrm{Tr}(\mathbf{P}_m \mathbf{Q} \mathbf{P}_m)$. Density matrices additionally provide a convenient way to describe quantum systems that have classical uncertainty. If a system is prepared in the state $\boldsymbol{\psi}_i$ with probability $p_i$, then the system is represented by the density matrix \cite{mikeike}
\begin{equation}
\mathbf{Q} = \sum_i p_i \boldsymbol{\psi}_i \boldsymbol{ \psi}_i^*.
\label{eqn:mixed-states}
\end{equation}
This example can be extended to the case in which the prepared state has a continuous distribution and ensures that measurement probabilities are properly modeled. A state with with a rank-one density matrix is known as a \textit{pure state} and corresponds to the case of no classical uncertainty about the prepared state. A state with a higher-rank density matrix is known as a \textit{mixed state} \cite{mikeike}.

 Systems of many qubits can be represented as extensions of a single qubit system with the tensor and Kronecker products. The state space of an $n$-qubit system is the tensor product of the $n$ component single qubit systems, i.e. $\mathbb{C}^{2^n} = \mathbb{C}^{2} \otimes\mathbb{C}^{2} \dots \otimes \mathbb{C}^{2}$. Any normalized vector in $\mathbb{C}^{2^n}$ is a valid state vector. For example, if $\mathbf{q}_1$ and $\mathbf{q}_2$ are single qubit systems, the two qubit composite system is given by $\mathbf{q}_1\otimes \mathbf{q}_2$. Analogously, if $\mathbf{Q}_1$ and $\mathbf{Q}_2$ are density matrix representations of two systems, the composite system has a density matrix of $\mathbf{Q}_1 \otimes \mathbf{Q}_2$. Any positive semidefinite operator with a trace equal to unity is a valid density operator. In multi-qubit systems, observables and measurements are simply defined in the relevant higher dimensional space. 

The \textit{fidelity} provides a notion of distance between quantum states \cite{mikeike}. The fidelity function $F(\mathbf{Q}_1, \mathbf{Q}_2) \in [0,1]$ is a symmetric function of its density matrix arguments. It is defined, for general mixed states as
$F(\mathbf{Q}_1, \mathbf{Q}_2) = \text{tr}((\mathbf{Q}_1^{\frac{1}{2}} \mathbf{Q}_2 \mathbf{Q}_1^{\frac{1}{2}})^{\frac{1}{2}})$ \cite{mikeike}. A low fidelity implies that states are ``far apart,'' and the fidelity is equal to unity if its arguments are the same. The fidelity between two pure states is $F(\mathbf{q}_1, \mathbf{q}_2)= |\mathbf{q}_1^* \mathbf{q}_2|$.  The fidelity between a pure state $\mathbf{q}$ and a mixed state $\mathbf{Q}$ is $F(\mathbf{q}, \mathbf{Q})= \sqrt{\mathbf{q}^* \mathbf{Q} \mathbf{q}}$. The fidelity can be used to induce a metric on states, $d$, via $d(\mathbf{Q}_1, \mathbf{Q}_2) = \arccos(F(\mathbf{Q}_1, \mathbf{Q}_2) )$ \cite{mikeike}.

\subsection{Stabilizer codes}

Stabilizer codes are a class of quantum error correcting codes designed to protect against a wide range of quantum errors \cite{gottesman1997stabilizer}. We briefly summarize their construction.

The $n$-qubit Pauli group, $\mathcal{P}_n$, is the set of operators in $\mathbb{C}^{2^{n}\times2^{n}}$ that can be written as a tensor product of $n$ of the $2\times2$ Pauli matrices $\mathbf{I}, \mathbf{X}, \mathbf{Y}, \mathbf{Z}$, up to a scalar multiple of $\alpha \in \{\pm 1, \pm j\}$, where
\begin{equation}
\mathbf{X} = \begin{pmatrix} 0 & 1 \\ 1 & 0 \end{pmatrix} \, , \;\;
\mathbf{Y} = \begin{pmatrix} 0 & -j \\ j & 0 \end{pmatrix} \, , \;\;
\mathbf{Z} = \begin{pmatrix} 1 & 0 \\ 0 & -1 \end{pmatrix} \; .
\end{equation}
The group multiplication operation is defined as standard matrix multiplication. Elements of $\mathcal{P}_n$ are unitary and are either Hermitian or skew-Hermitian. Thus, they are orthogonally diagonalizable, with eigenvalues $\pm 1$ or $\pm j$. Any two Pauli operators either commute or anti-commute.  

A stabilizer, $S$, is a commutative (Abelian) subgroup of $\mathcal{P}_n$ that does not contain the negative of the identity element, $-\mathbf{I}_{2^n}$. The additional requirement of closure under the group multiplication operation implies that elements of $S$ must have $\alpha = 1$. Thus, elements of $S$ are Hermitian operators with eigenvalues equal to $\pm 1$.

A \textit{stabilizer code} $C(S)$ is defined as the subspace of $\mathbb{C}^{2^n}$ formed by the intersection of $+1$ eigenspaces of the operators in $S$. An efficient description of the group $S$ is its generators. A set $G_S$ of generators of $S$ is a set of elements of $\mathcal{P}_n$ such that every element of $S$ is a product of elements in $G_S$. A generator set $G_S$ is called \textit{independent} if the set obtained by removing an element from $G_S$ fails to generate all elements of $S$. If $S$ is a stabilizer with an independent generator containing  $n-k$ elements, it can be shown that $C(S)$ is a $2^k$ dimensional vector space \cite{mikeike}. Furthermore, we also have that a state $\boldsymbol{\psi} \in C(S)$ if and only if $\mathbf{S}_n \boldsymbol{\psi} = \boldsymbol{\psi}$ for all $\mathbf{S}_n\in G_S$. 
Letting $s_i$ denote complex constants and $\vec{v}_i$ an orthonormal basis for $C(S)$, a general  \textit{codeword} for $C(S)$ can be written as
\begin{align}
\vec{x}= \sum\limits_{i = 0}^{2^{k}-1} s_i \vec{v}_i \; , \text{ with  } \sum\limits_{i = 0}^{2^{k}-1} |s_i|^2 = 1.
\end{align} A codeword is thus an arbitrary unit vector in $C(S)$.
%

There are several criteria that can be used to determine which quantum errors a stabilizer code can correct. 
A simple approach, which we adopt, is as follows. Consider a set of error operators $E\subset\mathcal{P}_n$. Each error $\mathbf{E}\in E$ either commutes or anticommutes with each generator of the stabilizer group. A sufficient condition for the stabilizer code to correct the errors in $E$ is for each $\mathbf{E}\in E$  to possess a unique commutation relationship with respect to the elements of $G_S$. Thus the stabilizer construction guarantees that each error $\mathbf{E}_k\in E$ maps the code space $C(G)$ bijectively to a $2^k$ dimensional subspace of $\mathbb{C}^{2^n}$. Furthermore, the uniqueness of the commutation relationships guarantees that different errors map $C(G)$ to different \textit{error subspaces} $\varepsilon_k$. Formally, $\varepsilon_k$ is the image of  $\mathbf{E}_k$ restricted to $C(G)$ 
(i.e.  $\varepsilon_k = \{ \textbf{y} \in \mathbb{C}^{2^n} | \text{ } \exists \text{ } \vec{x} \in C(G)\text{ with } \vec{y} = \mathbf{E}_k \vec{x} \}$)
and a unique commutation relationship guarantees that $\varepsilon_k \cap \varepsilon_j = \varnothing$ when $i\neq j$.

It should be stated that this criterion is sufficient but not necessary; the stabilizer formalism naturally lends itself to \textit{degenerate} quantum codes, where multiple errors yield the same syndrome and are correctable by the same operation. Consider a correctable error $\mathbf{E}$ and $\mathbf{V}\in S$. Both $\mathbf{E} \mathbf{V}$ and $\mathbf{E}$ will have the same commutation relations with respect to the stabilizer generators, and thus both $\mathbf{E} \mathbf{V}$ and $\mathbf{E}$ map an encoded state to the same subspace. Indeed, for $\vec{x} \in C(G)$ we have $\mathbf{EV} \vec{x} = \mathbf{E} \vec{x}$; namely, the effect of the errors on the codeword is exactly the same.

In the quantum setting, the stabilizer decoding process consists of performing projective measurements on the received state. The measurement observables are the stabilizer generators. This process projects the state into an intersection of the $+1$ or $-1$ eigenspaces of each $\mathbf{S}\in G(S)$. Thus, after the measurements the state collapses into one of the error subspaces $\varepsilon_k$.
The measurement  outcomes form a \textit{syndrome} (analogous to the classical syndrome) and identify into which subspace the state collapsed. The error correction conditions guarantee that the application of a correction (e.g., the error operator itself) for \text{any} correctable error yielding the measured syndrome recovers the encoded state \cite{gottesman1997stabilizer} \cite{mikeike}. This process demonstrates that a stabilizer code that can correct errors in a set $E$ can correct an arbitrary linear combination of correctable errors \cite{gottesman1997stabilizer}.

The projective measurements annihilate error operators that are not consistent with the measured commutation relationship. For example, consider the state $\vec{y} = (c_j\mathbf{E}_j + c_k\mathbf{E}_k)\vec{t}$ with $\vec{t}\in C(S)$. If $\mathbf{G}\in G_S$ anti-commutes with an error $\mathbf{E}_j$ but commutes with $\mathbf{E}_k$ and a measurement of $\mathbf{G}$ returns a $1$ (corresponding to a commutation),  the state after measurement is $\vec{\hat{y}} = (\mathbf{I}+\mathbf{G})\vec{y} = \mathbf{E}_k \vec{t}$

\section{System Model}

We consider a specific canonical received signal model for noncoherent wireless communication \cite{marzetta_capacity_1999} \cite{zheng2002communication}. The system has $N_{\text{TX}} = N_{\text{RX}} = 2$ antennas at both the transmitter and receiver. We assume a narrowband model with a single-tap MIMO channel $\mathbf{H}\in \mathbb{C}^{2\times 2}$. We assume a channel coherence time of $T=4$ channel uses. The transmitted, received, and additive noise signals are denoted by the complex $2 \times 4$ matrices $\mathbf{T}$, $\mathbf{Y}$, and $\mathbf{N}$, respectively, where the columns correspond to the time instants in the the coherence interval.  We take $\mathbf{N}$ to be a complex Gaussian random matrix with independent, identically distributed entries such that $[\mathbf{N}]_{i,j}\sim \mathcal{N}_{\text{C}}(0,\sigma_n^2)$. We further assume a Rayleigh fading model where the entries of $\mathbf{H}$ are independent and identically distributed with $[\mathbf{H}]_{i,j}\sim\mathcal{N}_{\text{C}}(0,1)$. Finally, we assume that $\mathbf{H}$ is constant over the coherence interval but that the channel realizations at different coherence intervals are independent. This model would be most appropriate for a frequency hopping system in an environment with rich scattering. The received signal over the coherence interval is now given by
\begin{equation}\label{eq:systemmodel}
    \mathbf{Y} = \mathbf{HT} + \mathbf{N}.
\end{equation}
Using the standard vectorization identity, letting $\mathbf{y} = \text{vec}(\mathbf{Y})$, $\mathbf{t} = \text{vec}(\mathbf{T})$, $\mathbf{n} = \text{vec}(\mathbf{N})$, and $\overline{\mathbf{H}} = \mathbf{I} \otimes \mathbf{I} \otimes \mathbf{H}$ we can write  (\ref{eq:systemmodel}) as
\begin{align}
\mathbf{y} = \overline{\mathbf{H}}\mathbf{t} + \mathbf{n} \; .
\end{align}
This particular form of the channel model is amenable to the design of a stabilizer code.

\section{Constructing a Noncoherent Space-Time Code via the Stabilizer Formalism}

In this section, we motivate the the application of quantum error correcting codes in a classical setting by observing that the communication channel at infinite SNR can be decomposed into a linear combination of Pauli group elements. 


The vectorized channel matrix $\overline{\mathbf{H}}$ highlights the coherence of the channel coefficients over time and admits a basis decomposition in the Pauli basis $\mathcal{P}_3$ of the form
\begin{equation}\label{eq:kronstru}
\begin{split}
    \overline{\mathbf{H}} &= \mathbf{I} \otimes \mathbf{I} \otimes ( c_0 \mathbf{I} + c_1 \mathbf{X} + c_2 \mathbf{Z} + c_3 \mathbf{Y} ) \\
     &= c_0 \mathbf{E}_0 + c_1 \mathbf{E}_1 + c_2 \mathbf{E}_2 + c_3 \mathbf{E}_3
\end{split}
\end{equation}
where 
\begin{subequations}\label{eq:channelcoeffs}  
    \begin{align}
        c_0 &= ([\mathbf{H}]_{1,1} + [\mathbf{H}]_{2,2})/2 \\
        c_1 &= ([\mathbf{H}]_{1,2}+[\mathbf{H}]_{2,1})/2 \\
        c_2 &= ([\mathbf{H}]_{1,1} - [\mathbf{H}]_{2,2})/2 \\
        c_3 &= j ([\mathbf{H}]_{1,2}-[\mathbf{H}]_{2,1})/2 
    \end{align}
\end{subequations}
and $\mathbf{E}_0 = \mathbf{I} \otimes \mathbf{I} \otimes \mathbf{I}$, $\mathbf{E}_1 = \mathbf{I} \otimes \mathbf{I} \otimes \mathbf{X}$, $\mathbf{E}_2 = \mathbf{I} \otimes \mathbf{I} \otimes \mathbf{Z}$, $\mathbf{E}_3 = \mathbf{I} \otimes \mathbf{I} \otimes \mathbf{Y}$. Defining $\mathbf{c} = [c_0, c_1, c_2, c_3]^T$ we have $\mathbf{c}\sim \mathcal{N}_{C}(0,\mathbf{I}_{4}/2)$.

The error set for the channel is thus $ E = \{ \mathbf{E}_0 , \mathbf{E}_1, \mathbf{E}_2, \mathbf{E}_3 \} $. This process is analogous to the quantum concept of channel discretization, in which a channel with a continuous set of possible realizations is equivalent to one that randomly applies a discrete set of error operators.


We now form a stabilizer group for this error set. The operators $\mathbf{S}_0 = \mathbf{X} \otimes \mathbf{Z} \otimes \mathbf{X}$ and $\mathbf{S}_1 = \mathbf{X} \otimes \mathbf{X} \otimes \mathbf{Z}$ satisfy the necessary commutation relations to form a set of stabilizer generators, as summarized in Table 1. 
\begin{table}[h]
\begin{center}
\begin{tabular}{ |M{2cm}||M{2cm}|M{2cm}|}
 \hline
 \multicolumn{3}{|c|}{Commutation Relationships} \\
 \hline

   & $\mathbf{S}_0$ & 
  
  $\mathbf{S}_1$ \\
  \hline \hline
    $\mathbf{E}_0$ & C & C \\
    
    $\mathbf{E}_1$ & C & A \\ 
    
    $\mathbf{E}_2$ & A & C\\
     
    $\mathbf{E}_3$ & A & A \\ 
 \hline
\end{tabular}
\end{center}
\caption{Table 1: Summary of Commutation Relations between stabilizer and error operators. C denotes commutation and A denotes anti-commutation}
 \vspace{-.5cm}
\end{table}

Because they commute, the stabilizer operators admit a partially intersecting $+1$ eigenspace, which has a two-dimensional basis spanned by the vectors 
\begin{subequations}
\begin{align}
    \mathbf{v}_0 &= 
    \begin{bmatrix}
    1 & 0 & 0 & -1 & 0 & 1 & 1 & 0
    \end{bmatrix}^T \\
    \mathbf{v}_1
    &=
    \begin{bmatrix}
    0 & -1 & -1 & 0 & -1 & 0 & 0 & 1
    \end{bmatrix}^T
\end{align}
\end{subequations}
We use these vectors to form a mapping that encodes two arbitrary complex numbers into a space-time code word. Given a  complex vector $\mathbf{s} = [s_1, \text{ }s_2]^T$ from a general codebook, we produce the vectorized space-time codeword by applying an encoding operator $\mathbf{C} =
\begin{bmatrix}
\mathbf{v}_0, & \mathbf{v}_1
\end{bmatrix}
\in \mathbb{C}^{8 \times 2}
$
giving 
\begin{equation}
    \label{vectorizedcodeword}
    \mathbf{t} =\mathbf{Cs}.
\end{equation} 
We assume that the symbol energy is normalized, i.e. $\mathbf{s}^*\mathbf{s} = 1$. This assumption coupled with the definition of $\mathbf{C}$ guarantees that $\mathbf{t}^*\mathbf{t} = 4$ which gives an average power of unity over the coherence interval.
The corresponding $2 \times 4$ code matrix for a codeword can be represented with the inverse vectorization operator $\text{vec}^{-1}:\mathbb{C}^{8} \mapsto \mathbb{C}^{2 \times 4}$,
\begin{equation}
\label{encoded}
\mathbf{T} = \text{vec}^{-1}(\mathbf{Cs}),
\end{equation}
or
\[
\mathbf{T}
=
\begin{bmatrix}
s_1 & -s_2 & -s_2 & s_1 \\ -s_2 & -s_1 & s_1 & s_2 
\end{bmatrix}.
\] This code is a \textit{generalized complex orthogonal design} and provides full diversity despite the noncoherent setting \cite{jafarkhani_space-time_2005}.

The symbol vector $\mathbf{s}$ is viewed as an information carrying qubit state, which we wish to preserve via the stabilizer encoding. Using the interpretation of a qubit as a 1-dimensional subspace of $\mathbb{C}^2$, we assume that symbol vectors $\mathbf{s}$ are drawn uniformly from a constellation $\mathcal{C}$. We choose our constellations as Grassmannian line packings in $\mathbb{C}^2$ \cite{strohmer2003grassmannian}. This choice is motivated in the following subsections.

\subsection{Decoding} 

In this setting of quantum-inspired classical coding, we can dispense with the ideas of quantum measurement and syndrome decoding in favor of the more familiar method of maximum likelihood (ML) inference. While our decoding process is based on computing the ML rule, it does lend itself to a quantum mechanical interpretation.

If we assume the encoded symbol $\mathbf{s}$ is drawn uniformly from some constellation $\mathcal{C}$, the maximum a posteriori rule reduces to the canonical ML problem of finding $\hat{\mathbf{s}}$ such that
\begin{align}
\hat{\mathbf{s}}=\arg\max_{\mathbf{s}\in\mathcal{C}} f_{\mathbf{s}|\mathbf{y}}(\mathbf{s}|\mathbf{y}) = \arg \max_{\mathbf{s}\in\mathcal{C}} f_{\mathbf{y}|\mathbf{s}}(\mathbf{y}|\mathbf{s}) \; .
\end{align} 

We begin by defining the following projection operators:
\begin{subequations}\label{eq:projects}
	\begin{align}
	\mathbf{P}_{0} &= (\mathbf{I}+\mathbf{S}_0)(\mathbf{I}+\mathbf{S}_1)/4 \\
	\mathbf{P}_{1} &=  (\mathbf{I}+\mathbf{S}_0)(\mathbf{I}-\mathbf{S}_1)/4 \\
	\mathbf{P}_{2} &=  (\mathbf{I}-\mathbf{S}_0)(\mathbf{I}+\mathbf{S}_1)/4 \\
	\mathbf{P}_{3} &= (\mathbf{I}-\mathbf{S}_0)(\mathbf{I}-\mathbf{S}_1)/4 \; .
	\end{align}
\end{subequations} 
Note that $\mathbf{P}_{0}$ is the projector onto the code space and $\mathbf{P}_{0} + \mathbf{P}_{1}  + 	\mathbf{P}_{2}  + \mathbf{P}_{3} = \mathbf{I}$ is the identity. The receiver computes the four corresponding projections of the received vector $\vec{y}$ onto the code space and the three error subspaces to obtain
\begin{subequations}\label{eq:ss}
	\begin{align}
	\mathbf{P}_{0} \vec{y} &= c_0 \vec{t} +  \mathbf{P}_{0} \vec{n} \\
	\mathbf{P}_{1} \vec{y} &= c_1 \mathbf{E}_1 \vec{t} +  \mathbf{P}_{1} \vec{n} \\
	\mathbf{P}_{2} \vec{y} &= c_2 \mathbf{E}_2 \vec{t} +  \mathbf{P}_{2} \vec{n} \\
	\mathbf{P}_{3} \vec{y} &= c_3 \mathbf{E}_3 \vec{t} +  \mathbf{P}_{3} \vec{n} \; ,
	\end{align}
\end{subequations} where the $c_0$, $c_1$, $c_2$, and $c_3$ are as defined in  (\ref{eq:channelcoeffs}) and we have used the fact that $\mathbf{P}_k \mathbf{E}_k = \mathbf{E}_k \mathbf{P}_0$. Since the projectors in  (\ref{eq:projects}) sum to identity, the vectors in  (\ref{eq:ss}) are sufficient statistics for $\mathbf{y}$. Recall that $\mathbf{c} = [c_0, c_1, c_2, c_3]^T \sim \mathcal{N}_{\text{C}}(\vec{0},\mathbf{I}/2)$ and is independent of the noise. 

The receiver now carriers out error correction on the projected vectors.  The receiver applies a unitary correction operator $\mathbf{E}_k$ to each projection $\mathbf{P}_k \vec{y}$ and obtains
\begin{subequations}
	\begin{align}
	\mathbf{z}_{0} &= c_0\vec{t} + \vec{n} \\
	\mathbf{z}_{1} &= c_1 \vec{t} +  \mathbf{E}_1 \mathbf{P}_1 \vec{n} \\
	\mathbf{z}_{2} &= c_2 \vec{t} +  \mathbf{E}_2 \mathbf{P}_2 \vec{n} \\
	\mathbf{z}_{3} &= c_3 \vec{t} +  \mathbf{E}_3 \mathbf{P}_3 \vec{n} \; .
	\end{align}
\end{subequations}

Since the $\mathbf{P}_k$ are orthogonal projection operators, the projected and corrected noise vectors, $\mathbf{E}_k \mathbf{P}_k \vec{n}$, are mutually independent. Following from this, the commutation relationships and unitarity of the correction operators imply that the resulting noise vectors are identically distributed with  $\mathbf{E}_k \mathbf{P}_k \vec{n} \sim\mathcal{N}_{\text{C}}(0,\sigma_n^2 \mathbf{P}_0 )$ for all $k$. Since $\vec{t} = \mathbf{C} \vec{s}$, projections of the $\mathbf{z}_k$ onto the column space of $\mathbf{C}$ are sufficient to estimate $\mathbf{s}$. Letting $\vec{n}_k = \mathbf{C}^* \mathbf{E}_k \mathbf{P}_k \vec{n}/(2\sqrt{2})$, and letting $\hat{c}_k = \sqrt{2} c_k$ the receiver computes
\begin{equation}
	\vec{q}_{k} = \dfrac{\mathbf{C}^* \vec{z}_{k}}{2\sqrt{2}}= \hat{c}_k \vec{s} +\mathbf{n}_k \; , \text{ for } k \in \{ 0, 1, 2, 3 \} \; .
\end{equation}

The $\mathbf{n}_k$ are independent and identically distributed with $\mathbf{n}_k\sim \mathcal{N}_{\text{C}}(\mathbf{0},\sigma^2_n\mathbf{I}/2)$. The scaled-identity covariance follows from the fact that $\mathbf{P}_0\mathbf{C} = \mathbf{C}$, since the columns of $\mathbf{C}$ are by definition in the code, and that $\mathbf{C}^*\mathbf{C} \propto \mathbf{I}$.

We now concatenate the $\mathbf{q}_k$ into the vector $\mathbf{q} = [\mathbf{q}_0^T\text{, } \mathbf{q}_1^T\text{, }\mathbf{q}_2^T\text{, }\mathbf{q}_3^T ]^T$ and reformulate our maximum likelihood problem as
\begin{align}\label{eq:map_ss_pdf}
\mathbf{\hat{s}} = \arg \max_{\mathbf{s}} f_{\mathbf{q}|\mathbf{s}}(\mathbf{q}|\mathbf{s}).
\end{align}  Given the transmit symbol $\mathbf{s}$, $\mathbf{q}$ is a Gaussian random vector. We define $\mathbf{w} = [\mathbf{\hat{c}}^T, \mathbf{n}_0^T, \mathbf{n}_1^T, \mathbf{n}_2^T, \mathbf{n}_3^T]^T$ so that $\mathbf{w} \sim \mathcal{N}_{\text{C}}(\mathbf{0},\boldsymbol{\Sigma})$, where
\begin{align}
\boldsymbol{\Sigma} &= \begin{bmatrix}
\mathbf{I}_{4} & \mathbf{0}_{4\times 8} \\ \mathbf{0}_{8\times 4} & \frac{\sigma_n^2}{2}\mathbf{I}_{8}
\end{bmatrix}.
\end{align} Defining the matrix $\mathbf{M}\in \mathbb{C}^{8 \times 12}$ via
\begin{align}
\mathbf{M} = \begin{bmatrix}
\left( \mathbf{I}_{4 } \otimes \mathbf{s} \right) & \mathbf{I}_{8}
\end{bmatrix},
\end{align} we have
\begin{align}
\mathbf{q} = \mathbf{Mw}.
\end{align} Thus, $\mathbf{q}\sim \mathcal{N}_{\text{C}}(\mathbf{0},\mathbf{Q})$, where $\mathbf{Q} = \mathbf{M\boldsymbol{\Sigma}M^*}$. It can be shown that
\begin{align}
\mathbf{Q} = \mathbf{I}_{4}\otimes \left(\mathbf{ss}^* + \frac{\sigma_n^2}{2}\mathbf{I}_{2\times 2}\right).
\end{align} It turns out that the second definition of $\mathbf{Q}$ is useful in simplifying the likelihood function. 

Assuming that $\mathbf{Q}$ is invertible, the likelihood function can be written as
\begin{align}\label{eq:pdf_gauss}
f_{\mathbf{q}| \mathbf{s}}(\mathbf{q}|\mathbf{s}) = \dfrac{\exp(-\mathbf{q}^*\mathbf{Q}^{-1}\mathbf{q})}{\pi^8  \text{det}(\mathbf{Q})}.
\end{align} Using the  property of determinants of Kronecker products yields $\text{det}(\mathbf{Q}) =  \text{det}(\mathbf{ss}^* + \frac{\sigma_n^2}{2}\mathbf{I}_{2\times 2})^4$. Since, by assumption $\mathbf{s}^*\mathbf{s} = 1$, we have $\text{det}(\mathbf{Q}) = [(1+\frac{\sigma_n^2}{2})\frac{\sigma_n^2}{2}]^4$, which is constant in $\mathbf{s}$. Furthermore, using the Kronecker product definition it is clear that $\mathbf{Q}^{-1}= \mathbf{I}_{4\times 4}\otimes (\mathbf{ss}^* + \frac{\sigma_n^2}{2}\mathbf{I}_{2\times 2})^{-1}$. Designating $\mathbf{U}_{\mathbf{s}} =(\mathbf{ss}^* + \frac{\sigma_n^2}{2}\mathbf{I}_{2\times 2})^{-1}$ and  calculating the inverse explicitly yields
\begin{align}\label{eq:inv_cov}
\mathbf{U}_{\mathbf{s}} = \dfrac{1}{\frac{\sigma_n^2}{2}(1+\frac{\sigma_n^2}{2})} \begin{bmatrix}
|s_2|^2+\frac{\sigma_n^2}{2} & -s_1s_2^* \\ -s_2s_1^* & |s_1|^2+\frac{\sigma_n^2}{2}
\end{bmatrix},
\end{align} which allows us to (finally) write down an explicit decision rule. Substituting (\ref{eq:pdf_gauss}) into (\ref{eq:map_ss_pdf}) and using the simplifications in (\ref{eq:inv_cov}) and the preceding paragraph motivate the decision rule
\begin{align}\label{eq:penMap}
\mathbf{\hat{s}} = \arg \min_{\mathbf{s}\in \mathcal{C}} \mathbf{q}^*(\mathbf{I}_{4\times 4}\otimes \mathbf{U}_{\mathbf{s}})\mathbf{q} \; .
\end{align}We simplify further by noting that, since $\mathbf{s}$ is normalized,
 \begin{align}\label{eq:redeaux_u}
\mathbf{U}_{\mathbf{s}} \propto \dfrac{\sigma^2}{2}\mathbf{I}_{2\time2} + (\mathbf{I}_{2\time2} - \mathbf{s}\mathbf{s}^*) \; .
 \end{align} Thus, using (\ref{eq:redeaux_u}),  (\ref{eq:penMap}) can be written
 \begin{align}\label{eq:map}
\mathbf{\hat{s}} = \arg \max_{\mathbf{s}\in \mathcal{C}} \sum\limits_{k = 0}^{3}\mathbf{q}_k^{*}\mathbf{s}\mathbf{s}^*\mathbf{q}_k = \arg \max_{\mathbf{s}\in \mathcal{C}} \mathbf{s}^*\sum\limits_{k = 0}^{3}(\mathbf{q}_{k}\mathbf{q}_k^{*})\mathbf{s} \; . 
 \end{align}
 
 This form of the decoding rule lends itself to a quantum mechanical interpretation. We interpret $\mathbf{\hat{q}}_k\mathbf{\hat{q}}_{k}^* = \mathbf{q}_k\mathbf{q}_k^*/\text{tr}(\mathbf{q}_k\mathbf{q}_k^*)$ as normalized density operators. We consider the mixed state, $\boldsymbol{\Psi}$  formed from drawing the states $\mathbf{\hat{q}}_k\mathbf{\hat{q}}_k^*$ with respective probabilities 
 \begin{align}
 p_k = \dfrac{\text{tr}(\mathbf{q}_k\mathbf{q}_k^*)}{\sum\limits_{i=0}^3\text{tr}(\mathbf{q}_i\mathbf{q}_i^*)} \; .
 \end{align} This yields the density matrix
 \begin{align}
\boldsymbol{\Psi} = \dfrac{\sum\limits_{i=0}^3\mathbf{q}_i\mathbf{q}_i^*}{\sum\limits_{i=0}^3\text{tr}(\mathbf{q}_i\mathbf{q}_i^*)}, \; 
 \end{align} which is the same matrix that appears on the right hand side of (\ref{eq:map}) up to a positive scale factor. Thus, using the definition of fidelity (cf. \ref{ss:intro2qm}) it can be seen that the ML detection rule consists of finding the input state that maximizes the fidelity with respect to $\boldsymbol{\Psi}$, or, more explicitly,
 \begin{align}\label{eq:castro}
\mathbf{\hat{s}} = \arg \max_{\mathbf{s}\in \mathcal{C}} F(\boldsymbol{\Psi},\mathbf{s}\mathbf{s}^*). 
\end{align}  We used the fact that maximizing the fidelity is the same as maximizing its square. We discuss our choice of constellation set $\mathcal{C}$ in the following section.

\subsection{Qubit symbol constellation}
The detection rule in  $(\ref{eq:map})$ motivates our choice of the Grassmannian frame for our qubit constellation. Consider the expectation
\begin{align}
\mathbf{B} = \mathbb{E}\left[\sum\limits_{k=0}^3\mathbf{q}_k\mathbf{q}_k^* \mid
\mathbf{s}\right] = 4 \mathbf{s}\mathbf{s}^*+2\sigma_n^2\mathbf{I}_2
\end{align} and consider the function $R_{\mathbf{s}}(\mathbf{\hat{s}}) = \mathbf{s}\mathbf{B}\mathbf{s}^*-\mathbf{\hat{s}}\mathbf{B}\mathbf{\hat{s}}^*$, where $\mathbf{s}\neq \mathbf{\hat{s}}$. This is the expected value of the difference between computing the statistic in  (\ref{eq:map}) on the transmitted symbol as opposed to another, not transmitted symbol. We expect that the dominant error will occur when $R_{\mathbf{s}}(\mathbf{\hat{s}})$ is minimized over all $\mathbf{s}$ and $\mathbf{\hat{s}}$. We therefore seek a constellation set with the maximal minimum $R_{\mathbf{s}}(\mathbf{\hat{s}})$. Since the transmit symbols are normalized, the definition of $R_{\mathbf{s}}(\mathbf{\hat{s}})$ indicates that for a $N$-point constellation encoding $\log_2(N)$ bits we should select the set given by
\begin{align}
    \hat{\mathcal{C}} = \min_{\mathcal{C} = \{\mathbf{s}\in \mathbb{C}^2 \mid \mathbf{s}^*\mathbf{s} = 1  \}, |\mathcal{C}| = N} \max |\mathbf{\hat{s}}^*\mathbf{s}|^2
\end{align}
This indicates that we should choose our constellation as a Grassmanian frame \cite{strohmer2003grassmannian} \cite{loveHeath}. Furthermore, in the quantum picture, this is akin to choosing input states that are maximally far apart with respect to a metric induced by fidelity.

\section{Simulation Results and Conclusions}
 In this section, we present simulation results to demonstrate the performance of the noncoherent space-time code presented in Section IV. We have considered Grassmannian packings of size $N = 4$ and $N = 8$ in $\mathbb{C}^2$ with a Rayleigh fading environment. Specifically, we used the Grassmanian packings listed on \cite{lovepacks}. 

We compared the stabilizer-based, non-coherent construction with a coherent scheme based on the Alamouti code (for $2\times2$ systems) at spectral efficiency rates of $r =1/2$ and $r=1$ bits/channel use \cite{bigAl}. Channel estimation is first performed by transmitting the symbols $[1,1]^T/\sqrt{2}$ and $[1,-1]^T/\sqrt{2}$ and solving for an estimate of $\mathbf{H}$ at the receiver. We then use the Alamouti scheme to transmit one space time symbol $\mathbf{s}\in \mathbb{C}^2$ over the remaining two channel uses in the coherence interval. We encode in $\mathbf{s}$ two binary phase-shift keying (BPSK) symbols for the rate $r=1/2$ approach and two quadrature phase-shift keying (QPSK) symbols for the rate $r=1$ approach. 

Similarly, we compare to an approach using differential unitary group codes, as outlined in \cite{differential}. At both $r=1/2$ and $r=1$, the first two channel uses are used for the $2 \times 2$ reference matrix, and no information is transmitted. With the next two channel uses we transmit a single differentially encoded $2 \times 2$ matrix drawn from an appropriately sized constellation. For $r=1/2$, this constellation is a group code over the QPSK constellation; specifically, we encode over the $2 \times 2$ Pauli group elements. For $r = 1$, this constellation is a dicylic group code generated over $16$-PSK. The transmit symbols in both sets of comparisons are appropriately normalized so that the transmit power is constant over the four transmissions.

\begin{figure}[h]
	\centering
	\includegraphics[scale = .57]{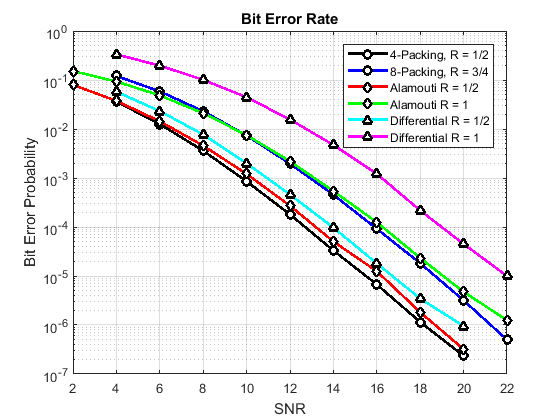}
    \vspace{-.3cm}
	\caption{Bit error rate for various packings. We simulated 10 million channel realizations (assumed to be coherent for four instances each). For SNR's with bit error rates lower than $10^{-6}$, we simulated 100 million channel realizations.}
	\label{fig:res2}
\end{figure}

\section*{Acknowledgments}
This work was supported by the U.S. Office of Naval Research under Grant No.\ N00014-17-1-2107.  T.C. received additional support from an Engineering Doctoral Fellowship given by the University of Texas at Austin Cockrell School of Engineering.



%
\bibliographystyle{IEEEtran}
\bibliography{IEEEabrv,refs} 

\end{document}